# Towards Understanding of Gold Interaction with AIII-BV Semiconductors at Atomic Level

B.R. Jany,[a]* A. Janas,[a] W. Piskorz,[c] K. Szajna,[a] A. Kryshtal,[b] G. Cempura,[b] P. Indyka,[c,d] A. Kruk,[b] A. Czyrska-Filemonowicz,[b] F. Krok.[a]

[a]The Marian Smoluchowski Institute of Physics, Jagiellonian University, Lojasiewicza 11, 30-348 Krakow, Poland
[b]The International Centre of Electron Microscopy for Materials Science, AGH University of Science and Technology, 30-059 Krakow, Poland
[c]The Faculty of Chemistry, Jagiellonian University, ul. Gronostajowa 2, 30-387 Krakow, Poland
[d]Malopolska Centre of Biotechnology (MCB), Jagiellonian University, Krakow, 30-387, Poland

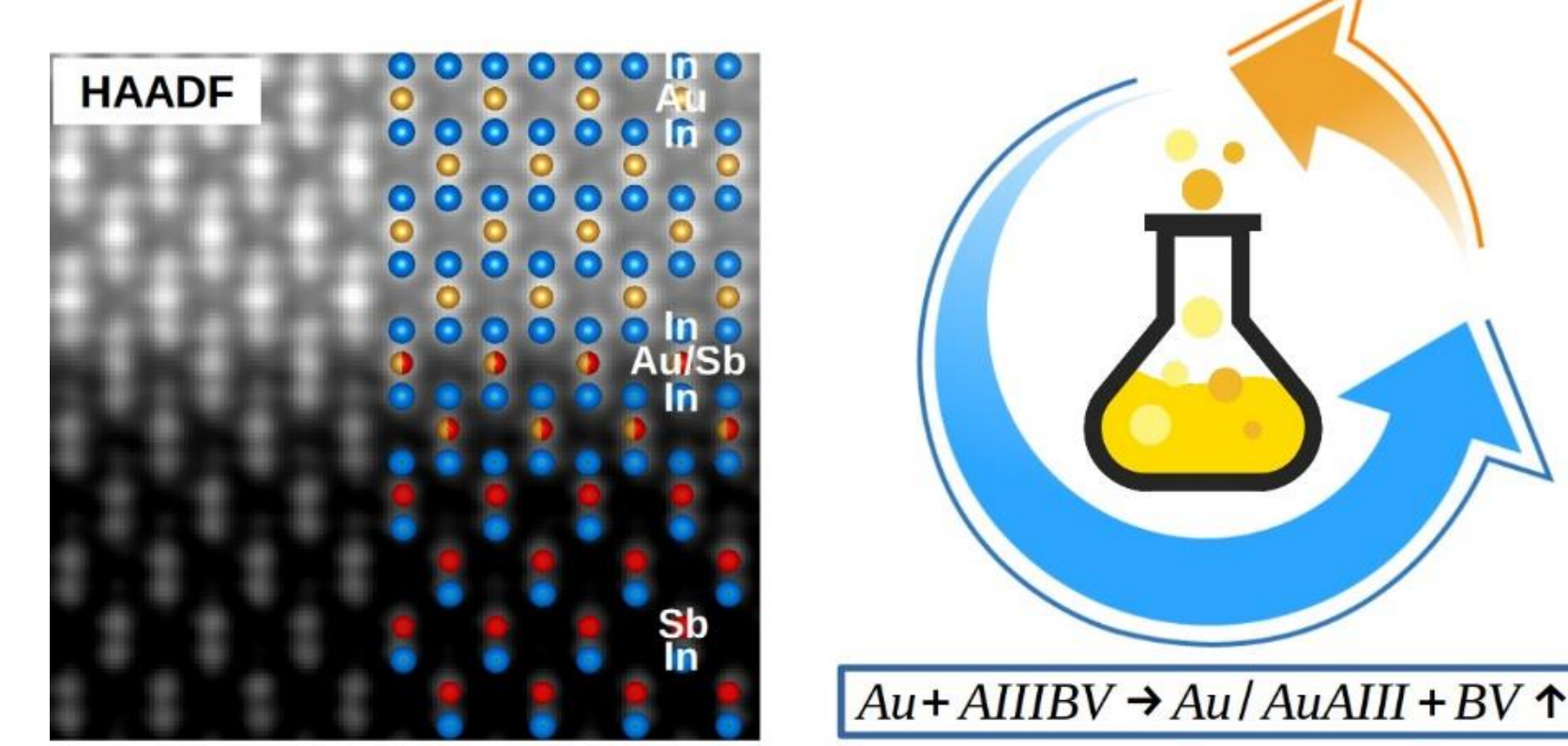

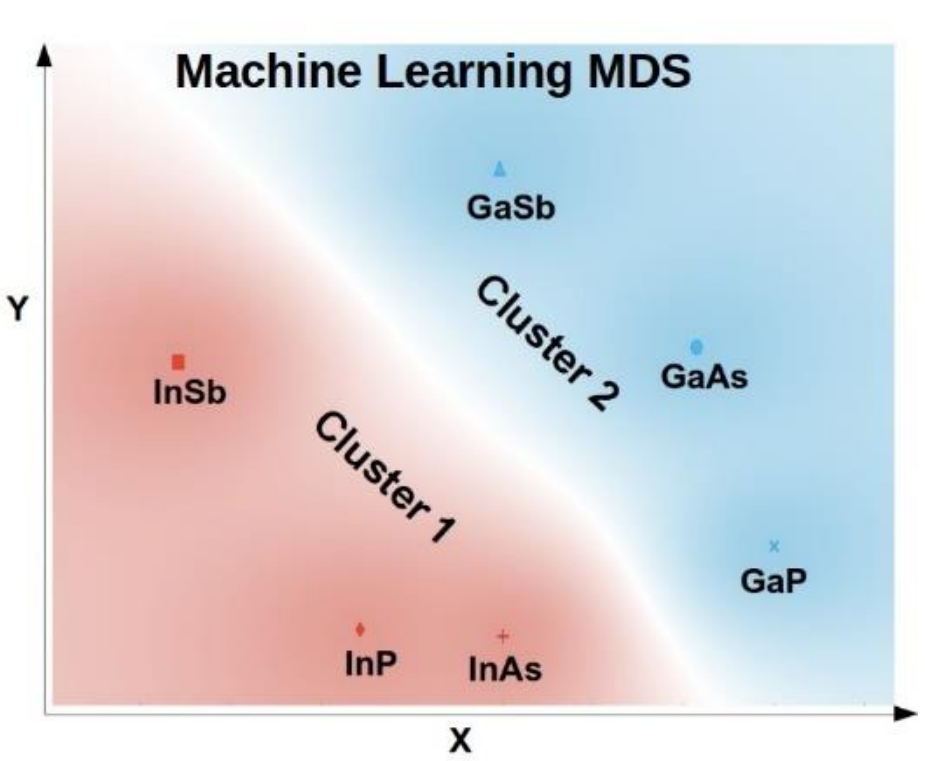

## Abstract

AIII-BV semiconductors have been considered for decades to be a promising material in overcoming the limitations of silicon semiconductor devices. One of the important aspects within AIII-BV semiconductor technology are gold-semiconductor interactions on the nanoscale. We report on investigations into the basic chemical interactions of Au atoms with AIII-BV semiconductor crystals by an investigation of nanostructures formation in the process of thermally-induced Au self-assembly on various AIII-BV surfaces, and this by means of atomically resolved High Angle Annular Dark Field (HAADF) Scanning Transmission Electron Microscopy (STEM) measurements. We have found that the formation of nanostructures is a consequence of the surface diffusion and nucleation of adatoms produced by Au induced chemical reactions on AIII-BV semiconductor surfaces. Only for InSb crystal we have found that there is efficient diffusion of Au atoms into the bulk, which we experimentally studied by Machine Learning HAADF STEM image quantification and theoretically by Density Functional Theory (DFT) calculations with the inclusion of finite temperature effects. Furthermore, the effective number of Au atoms needed to release one AIII metal atom has been estimated. The experimental finding reveals a difference in the Au interactions with In- and Ga-based groups of AIII-BV semiconductors. Our comprehensive and systematic studies uncover details of the Au interactions with the AIII-BV surface at the atomic level with chemical sensitivity and shed new light on the fundamental Au/AIII-BV interactions at the atomic scale.

*Corresponding author e-mail: benedykt.jany@uj.edu.pl

AIII-BV semiconductors due to their unique properties, such as high electron mobility and direct bandgap, are being considered material for the new generation of nanoscale electronic devices.[1–4] Recently, new technologies were developed such as template-assisted selective epitaxy (TASE) by IBM Zurich group[5] and epitaxial lift off (ELO) technique[6] to integrate AIII-BV with Si at the nanoscale. This will allow one to expand the use of AIII-BV crystals by extending the conventional Si-based technology for future AIII-BV/Si nanodevice fabrication.[7] Furthermore, for many applications arrays of standing, vertically aligned AIII-BV nanowires grown on semiconductor surfaces are desired.[8–10]Among others, a device for efficient water splitting can be built from such AIII-BV nanowires arrays.[11,12] Also the AIII-BV nanowire LED device integrated on Si has been developed, with 3 orders of magnitude higher efficiency than the conventional device.[13] To control the physico-chemical properties of the devices local information on the atomic arrangement is necessary as has recently been shown by atomic resolution spectrum imaging in STEM.[14] The growth of such monocrystalline AIII-BV nanowires is mainly catalyzed by Au seeded nanoparticles.[15] Studies on the role of Au in the process of nanowires growth, in a closed system being in equilibrium, have been performed by Dick *et al.*[16] They have shown that deposited Au is not inert with respect to AIII-BV material and interacts with the substrate forming a variety of intermetallic compounds (alloys) within the Au-AIII system. In this respect the process of the thermally controlled self-assembling of an Au thin layer on reconstructed AIII-BV surfaces in UHV conditions is a method of choice to create patterns of well-ordered nanostructures. Recently it has been shown that the thermally activated motion of Au droplets on the AIII-BV crystal surfaces can lead to new asymmetric morphologies for use in nanophotonic devices applications.[25] The interactions of Au with AIII-BV semiconductors and the development of new phases is governed by the relations of Au-In-BV and Au-Ga-BV ternary phase diagrams[17–19] (see also the supporting information in Fig. S2). These diagrams show that there are varieties of stable phases in the Au-AIII systems (see also Table S2) and only two stable phases for Au-BV pairs, i.e., $AuSb_2$ and $Au_2P_3$. Thus, the interactions of Au with the whole group of AIII-BV are dominated by the Au-AIII phases formation as defined by the Au-In and Au-Ga binary phase diagrams.[20,21] The above rules hold also in the present case for "an open system", where the BV element is volatile and escapes to the vacuum while annealing the samples.[22,23]

Taking into account the great role of AIIIBV semiconductors in the construction of new systems, and on the other hand, the requirements of their progressive miniaturization, it is very important to study the semiconductor-metal interfaces at the atomic level. In the present paper, we explore the interactions of Au with AIII-BV semiconductors at an atomic scale not accessible before, via thermally induced self-assembly of 2 Mono-Layers (ML) of Au on various AIII-BV (001) semiconductor crystal surfaces. We have used state-of-the-art techniques, as available today. We prepared all the samples in ultra-high vacuum (UHV) conditions (pressure of about $10^{-10}$ mbar) to prevent any contaminations. All the samples were later imaged with atomically resolved High Angle Annular Dark Field (HAADF) Scanning Transmission Electron Microscopy STEM, which delivers a true chemical contrast at atomic resolution. Finally, the collected data was processed by Machine Learning based multivariate statistical analysis. For the diffusion data interpretation, we used Density Functional Theory (DFT) with the inclusion of temperature effects via the harmonic vibrational analysis to obtain the details of the thermally-induced diffusion. Our research shed new light on the fundamental Au/AIII-BV interactions, i.e., the effectiveness of bond breaking at the atomic scale and diffusion at the atomic scale. This fundamental knowledge is required for the tailored applications in the field of the manufacturing of AIII-BV semiconductors devices. We have found that there is a large difference between the Au interactions for In- and Ga-based substrates. Our investigations prove that the Au content in the nanostructures formed depends on the binding energy of the AIII-BV components. The formed nanostructures could be used as seeds for the nanowires growth, substituting Au seeding by nanoparticles.[24] The formed metallic nanostructures of different stoichiometry influences the electronic properties of metal-semiconductor nanodevice due to the appearance of the Schottky barrier (the Schottky contact is formed). When the barrier value approaches to zero, the ohmic contacts are formed.[26,27] The electronic properties of the interfaces formed have impact on the development of future nanoelectronic devices based on AIII-BV technology,[28] such as nano Schottky diodes.[29,30] Finally, we uncovered that the self-assembly process driven by chemical reactions of Au with the substrate (effectiveness of bond breaking) is the main force of the nanostructures growth on the sample surfaces.

# Results

## In-based AIII-BV semiconductor substrates - InSb(001), InAs(001) and InP(001)

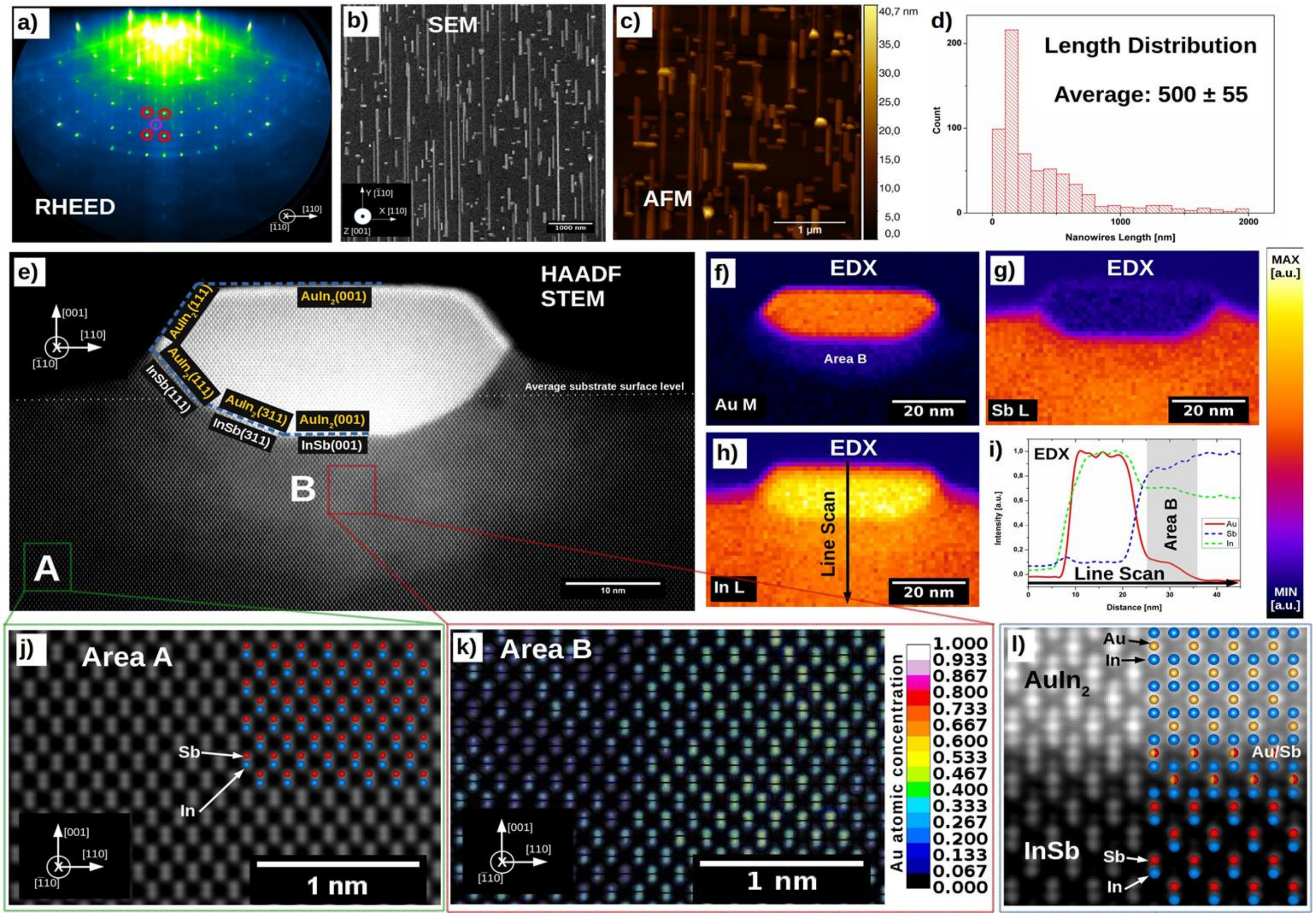

*Figure 1: Properties of the metallic nanowires grown on InSb(001) surface after the deposition of 2ML of Au at 330 °C. In a) the RHEED pattern of the clean and reconstructed c(8x2) InSb(001) surface is shown. In b) SEM and in c) AFM morphology of the nanowires, with their length distribution in d), are presented. In e) an atomically resolved HAADF STEM image of the nanowire cross section is depicted. In the image, the region of bright contrast just below the nanowire (marked as area B) corresponds to the area enriched in Au. The chemical composition of the nanowire and the surroundings is provided with EDX maps of Au (in f), Sb (in g) and In (in h) and also with the EDX line profile through the nanowire/substrate interface (seen in i). An atomically resolved HAADF image of the pure InSb substrate (of the area marked with A), partially overlapped with the InSb atomic structural model, is shown in j). In k) an Au atomic concentration map, i.e., the amount of Au incorporated into InSb lattice, of the region B is presented. The value "1" corresponds to the Au concentration of 100 at. %. In l) the HAADF image of the interface between the $AuIn_2(001)$ nanowire and InSb(001) substrate surface, together with the proposed atomistic structural model of the interface, is shown.*

The deposition of 2ML of Au on a clean c(8x2) reconstructed surface of InSb(001) monocrystal (see Fig. 1a) at 330 °C results in the formation of long nanowires oriented along the [-110] substrate surface direction as depicted in Fig. 1b). The nanowires have a length of up to few micrometers, average width of ~70 nm and a constant height of ~5 nm as derived from the Scanning Electron Microscopy (SEM) and Atomic Force Microscopy (AFM) measurements (Fig.1c and d).

In Fig. 1e the atomically resolved High Angle Annular Dark Field (HAADF) Scanning Transmission Electron Microscopy (STEM) image of the nanowire's cross section is presented. In the HAADF STEM technique the imaging contrast is proportional to the atomic number Z and, thus, this method ensures the chemical sensitivity (the Au atomic columns appear brighter than in the In and Sb columns). Due to the very high sensitivity of this technique, small variations in the composition for the AIII-BV alloys could be detected.[31] It is seen that the nanowire is crystalline and is composed of Au and In without Sb as indicated by the EDX measurements (Fig.1f-i). This points out that during the process of nanowire growth, the Au interactions with InSb substrate surface result in breaking of the bonds between In and Sb atoms. A stable Au-In phase is formed (see the supporting information in Fig. S2a and Table S2) while Sb atoms are emitted to vacuum. By examining the interplanar spacing of the atomic columns and their contrast in the HAADF STEM imaging, the nanowire composition is identified as an $AuIn_2$ alloy.[21] $AuIn_2$ nanowire's stoichiometry is also confirmed by the SEM/EDX measurement results which were analyzed by a newly developed method of chemical quantification based on the Machine Learning approach.[32] The driving process of the chemical reaction under consideration on the sample surface has been found to be: $Au + 2InSb \rightarrow AuIn_2 + 2Sb\uparrow$ ($\Delta H = -0.215$eV).[33,34] From the reaction stoichiometry, one can conclude that for these experimental conditions one Au atom releases two AIII metallic atoms (here In) on the InSb (001) substrate surface while BV (here Sb) atoms are released to the vacuum. In Fig. 1l an atomically resolved HAADF STEM image of the interface between $AuIn_2$ (001) and InSb (001) is presented. The epitaxial relations of the interface are clearly demonstrated by the continuation of the atomic rows between the $AuIn_2$ nanowire and InSb substrate. In Fig. 1e there are marked that $AuIn_2$ nanowires grow epitaxially on the InSb substrate with the epitaxial relationship of (001)$AuIn_2$//(001)InSb and [110]$AuIn_2$//[110]InSb. A proposed structural model of the interface is overlaid on the atomically resolved HAADF STEM image (Fig. 1l). Such epitaxial relations of the system were already predicted by the computational model.[34,35] The nanowire is partially buried below the average substrate surface level and is surrounded on its sides by a crystalline rim having a common (111) facet plane. The $AuIn_2$ nanowire exposes on its top the (001) plane of square-symmetry in agreement with previous atomically resolved imaging with non-contact atomic force microscopy.[36] The formation of the $AuIn_2$ phase after annealing of Au layer on InSb(111) was previously observed by HRTEM[38] but without chemical sensitivity at the atomic level.

From Fig. 1e it is also clearly seen that the HAADF STEM contrast of the InSb substrate crystal is not uniform. The region just below the nanowire (marked as the region B) is brighter than the rest of the InSb bulk (region A) indicating that the region B could be enriched in Au due to the diffusion (dissolution) of Au atoms into the InSb crystal. Indeed, the corresponding EDX measurements (Fig. 1f-i) exhibit that the region B is enriched in Au and there is depletion of Sb. To distinguish between pure In-Sb columns and those containing Au atoms (assuming constant sample thickness) the HAADF image quantification was performed. Due to the noise inherent to the experiment, intermixing between the Au and In-Sb phases is studied statistically using Machine Learning algorithms (Random Forest) as implemented in Trainable Weka Segmentation[37] (for details see the supporting information in Fig. S1 and Table S1 as well as the Method section). The results of Machine Learning HAADF image quantification of the Au enriched region is presented in Fig. 1k, as the Au atomic concentration map. We have found that Au does not uniformly populate the In-Sb lattice atomic positions (see the supporting information in Fig. S2d, Table S1).

The observed Au diffusion into InSb lattice was still investigated in detail and the results are shown in Fig. 2. Fig. 2a shows the atomically resolved HAADF STEM of one of the In-Sb dumbbells from Fig. 1j (for better visibility of Z contrast a false colouring was used). One can clearly distinguish between the A(In) and B(Sb) atomic columns. The Au atomic concentration maps in: antimony atomic sub-lattice A(Sb) (see Fig. 2b) and indium atomic sub-lattice B(In), (see Fig. 2c) were extracted from the results of the atomic HAADF image quantification. It has been found that the antimony sub-lattice is more populated with Au (on average ~18 at. % of Au) than the indium atomic sub-lattice (on average ~13 at. % of Au). Furthermore, the Au concentration in both sub-lattices decreases when the distance from the island increases. To characterize the difference in Au incorporation into the substrate atomic sub-lattices the map of atomic concentration ratio R was constructed, defined as: R=[Au in A(Sb)]/[Au in B(In)], and shown in Fig. 2d. In Fig. 2e a plot of the R ratio value against the average Au concentration in the paired In-Sb atomic columns is presented, where two regions of high and low R ratio are clearly visible. Thus, for the low Au concentration, the Sb lattice is much more populated with Au than the In lattice ($R \gg 1$) while for the high Au concentrations both sub-lattices are equally populated with Au (R ratio is close to 1).

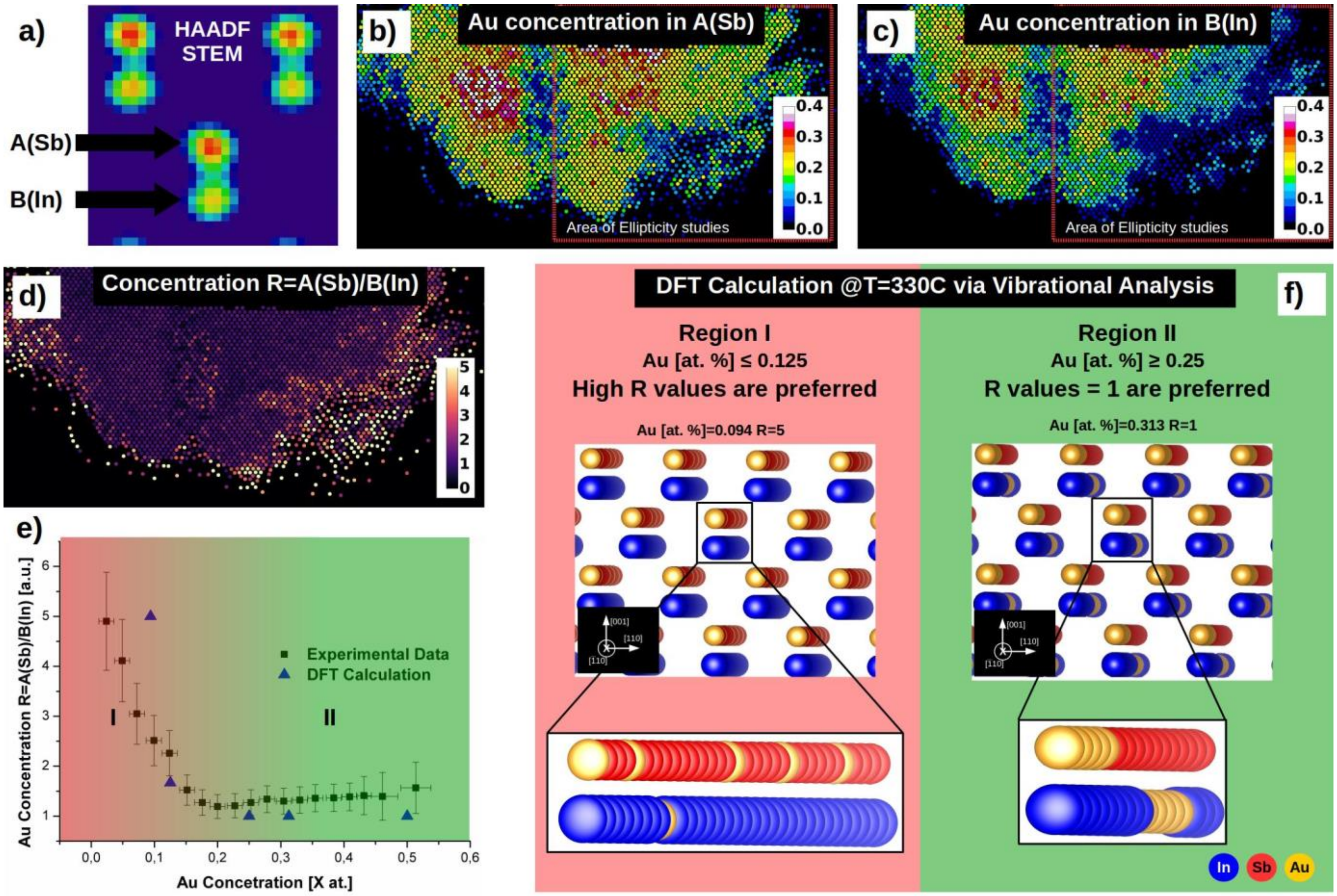

*Figure 2: Analyses of the InSb substrate region enriched with Au [corresponding to the region B in Fig. 1e)]. In a) false coloring atomically resolved the HAADF STEM of antimony A(Sb) and the indium B(Sb) atomic columns. In b) and c) maps of Au concentration in antimony sub-lattice A(Sb) and in indium sub-lattice B(In), resulting from the Machine Learning HAADF STEM image quantification, are shown, respectively. In d) the map of Au concentration ratio R=[Au in A(Sb)]/[Au in B(In)] is presented. In e) the Au concentration ratio R as a function of the average Au concentration in particular atomic column (value 1 corresponds to 100 at. %) is shown. Two regions, called I and II, are clearly visible - for the small Au concentrations (below ~20 at. %) the Au preferentially builds up into the antimony A(Sb) lattice resulting in R>>1. As the Au concentration increases (~20 at. % and above), the population in Au of both lattices becomes equal (R~1). In f) there are results of DFT calculations at T=330 ºC showing stable Au/InSb structures for different Au concentration. The calculated ratio R is presented as blue triangles together with experimental data in e).*

Since the HAADF STEM measurements show only mean concentration of Au in particular columns of In and Sb, without the possibility to determine the distribution of dissolved Au atoms in the column, DFT simulations were performed. We employed state of the art first principles DFT calculations with the inclusion of the finite temperature effect *via* the harmonic vibrational analysis (T=330 ºC) yielding the vibrational entropy contribution to the Gibbs free energy. The purpose of these simulations was to determine whether Au atoms are uniformly distributed in the columns or they tend to form compact clusters. In total, we tested about 50 different model configurations of the Au incorporation into the InSb lattice, from which we selected those with the lowest Gibbs free energy, accordingly to the probability of structure realization as the Boltzmann Distribution. DFT simulations, supporting the experimental results, distinguish two basic configurations depending on the concentration of Au in the InSb lattice. For Au concentrations less than 20 at. %, Au atoms are preferably incorporated into the Sb sub-lattice, providing high R ratio. In this case DFT results indicate that Au atoms are homogeneously distributed in atomic columns, not leading to formation of any Au agglomerates. On the other hand, for higher Au concentrations, above 20 at. %, Au is equally incorporated in both sub-lattices (R ratio is about 1). In this case, however, Au atoms agglomerate leading to the formation of Au dimers/oligomers in individual columns as the most energy preferred configurations (having the lowest free Gibbs energies). Thus, the formation of Au atomic wire-like structures in the InSb lattice is predicted. Examples of configurations of the arrangement of Au atoms in the InSb lattice, for both regions, are shown in Fig. 2f.

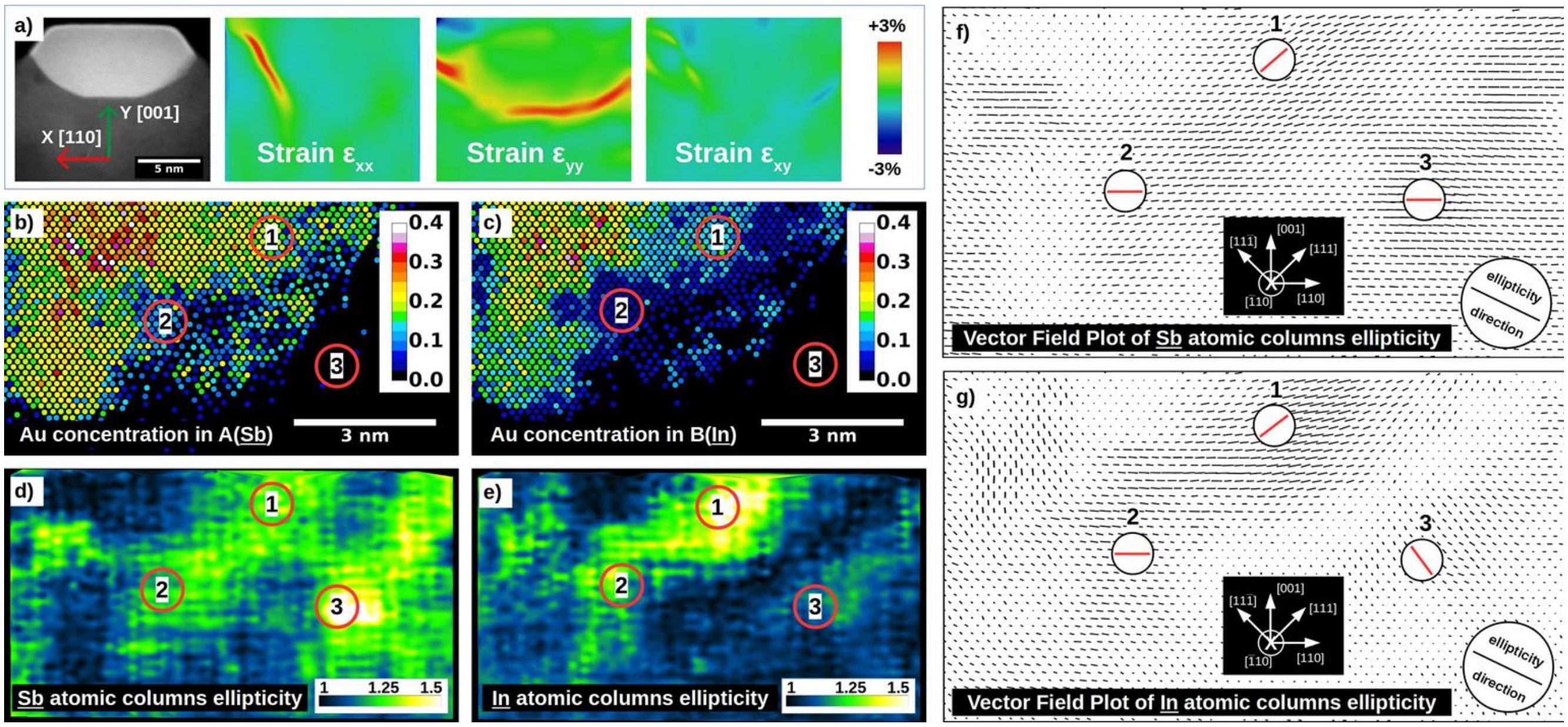

*Figure 3: Analyses of the structural properties of the metallic nanowire/InSb substrate interface. In (a) maps of strain resulted from the Geometric Phase Analysis (GPA) are presented. Just below the nanowire (in the area of ~2nm thickness) a positive ~3% εyy strain field is observed. There is no correlation observed between the GPA lattice strain and the amount of Au in the InSb lattice. There is no GPA strain visible in the Au diffusion area (area of ~10 nm below the nanowire). In (b) the Au concentration maps for antimony A(Sb) sublattice and indium B(In) sublattice (in c) with corresponding atomic columns ellipticity maps for antimony (in d) and indium (in e) are shown. In (f) the Vector Field Plots of antimony atomic columns ellipticity and indium (in g) atomic columns ellipticity are presented. The maps are prepared for the sample areas presented in Fig. 2b-c. The length of the vector line is proportional to the ellipticity value. Three different regions, called 1, 2 and 3, are marked on the maps (see discussion in the text).*

The main outcome of the performed DFT calculations is the detailed determination of the structures at the atomic level, which is not accessible experimentally by HAADF STEM due to the integration of the signal within particular atomic columns along the sample (lamella) thickness. The agreement between the DFT and experimental results is due to the inclusion in the DFT simulations the thermo-vibration effects, which predict the correct energy dependence between the studied structural atomic models. The analysis of solely SCF electronic energy (without temperature effects), contradictory to the experiment, predicts the preferential replacement of indium upon Au diffusion into atomic lattice of In-Sb. So this shows the importance of temperature effects for the diffusion studies.

Using the Geometric Phase Analysis (GPA) we also explored the effect of Au inclusion on the InSb crystal lattice strain. The lattice strain, obtained by GPA from the analyses of the atomically resolved HAADF STEM images, traces the local deformation of an atomic lattice unit cell in the x-y plane perpendicular to the electron beam (z direction). The results are presented in Fig. 3a. Just below the nanowire, in the region of around 2 nm, is visible a positive ~3% of $\varepsilon_{yy}$ growth strain field, representing a ~3% expansion of the atomic lattice in the y direction (growth direction). No correlation between the observed strain fields and the amount of Au atoms incorporation into the InSb lattice has been found, since the same strain field goes through the area of different Au concentration. According to the experimental data, below the ~2nm strain field, in the region where the Au diffusion still takes place (the area of ~10 nm below the wire) is strain free. This observation is consistent with our DFT simulations prediction in the sense that upon the Au incorporation two contradictory effects are observed. On one hand, the lattice swelling due to the incorporation of Au atoms. The DFT simulations showed that the resulting Au-Sb and Au-In bonds are longer (2.834 Å and 2.820 Å, respectively) than the DFT calculated length of the In-Sb bond (2.805 Å). And, on the other hand, as calculations show when the equilibrium Au doping takes place the Au-dimers are formed of short Au-Au bonds (2.799 Å) that contribute to the shrinkage of the lattice. Thus, no significant lattice swelling is observed as these two effects cancel out each other to a significant extent (*vide supra*).

Furthermore, to elucidate the In and Sb atoms off-axis positioning within the atomic columns due to the Au atoms incorporation, we performed the atomic columns shape analysis *via* ellipticity analysis of the HAADF STEM images.[39,40] Each of the atomic columns was fitted with an elliptical 2-D Gaussian. The ellipticity parameter was extracted and defined as the ratio of a long-to-short ellipse axis accordingly to Nord *et al.*[40] The ellipticity value of “1” corresponds to perfectly circular atomic columns while values greater than “1” indicates displacement of the atoms, within the column, in the x-y plane.[39,40] The ellipticity maps, corresponding to Au concentration maps of Fig. 2b-c, are plotted in Fig. 3d-e for the Sb and In atomic columns, respectively. In the regions where indium and antimony lattices are both highly populated with Au the ellipticity is close to unity. In Fig. 3f-g the vector field plots show the directionality of the atoms displacement in the x-y plane (the length of the vector is proportional to the ellipticity value). It is clear that the Au concentration in the InSb crystal highly influences the ellipticity of atomic columns. Thus, for three regions marked as (1), (2), (3) in Fig. 3b-g it can be stated:

- (1) for Au concentration in In sub-lattice of ~11 at. % and in Sb sub-lattice of ~18 at. % (R value ~ 1.6), the ellipticity for indium and antimony atomic columns are about 1.4 and 1.2, respectively. The ellipticity direction is the same for both sub-lattices i.e., the atomic displacement within the column are happening in the InSb [111] bulk direction;
- (2) for the case of an Au concentration in In sub-lattice of of ~5 at. % and in Sb sub-lattice of ~11 at. % (R value ~2.2), the ellipticity values (of about 1.2) and ellipticity direction are comparable for both sub-lattices;
- (3) for the case of the Au concentration in Sb sub-lattice of ~5 at. % and no diffusion into the In sub-lattice (very high R value), the ellipticity for Sb is high (~1.4) and directed along [110] direction of the InSb crystal, while the ellipticity for In is smaller (~1.1) and is directed along [111] InSb direction.

In Fig. 4 a)-b) the SEM and AFM morphologies of surface nanostructures resulted from the deposition of 2 ML of Au on InAs(001) and InP(001) at 330°C are shown, respectively. In both cases, the nanostructures are less anisotropic than the ones formed for the Au/InSb(001) system. In the case of Au/InAs(001), the nanostructures are slightly elongated along the [-110] direction and have an average size of 36 nm whereas for Au/InP(001) round nanostructures with sizes of 31 nm are formed. The atomically resolved HAADF images (Fig. 4 c-d) show that for both cases the nanostructures are in registry with the substrates. It has been found that for the case of Au/InAs(001) system the developed nanostructures are composed of an $Au_3In$ alloy with the epitaxial relationship (001)$Au_3In$//(001)InAs and [100]$Au_3In$//[110]InAs. We identify the driving process of the chemical reaction for this system as: $3Au + InAs \rightarrow InAu_3 + As\uparrow$ ($\Delta H = -0.007$eV).[33,34] In this case the nanostructures grow above the average substrate surface level as indicated in Fig. 4c. This is related to the kinetics of the growth process where Au atoms break the bonds of InAs substrate dimers and nucleate with the released In atoms to form finally clusters (islands) of In/Au alloy (As atoms evaporate into the vacuum). The islands form a kind of mask, which blocks further InAs bond breaking underneath. Thus, further incoming Au atoms can break only InAs bonds which are outside of the mask region inducing the lowering of the substrate surface level outside of the already developed $Au_3In$ nanostructures. The detailed view on the interface with the atomistic structural model is presented in Fig. 4d.

In turn, for the Au/InP(001) system the nanostructures formed are of the $AuIn_2$ phase (see Fig. 4e-f) with the epitaxial relationship (-11-1)$AuIn_2$//(001)InP and [110]$AuIn_2$//(110)InP. This defines the chemical reaction as $\mathrm{Au+2InP} \rightarrow \mathrm{AuIn_2} + 2P \uparrow$ (ΔH=0.172eV).[33,34] The formed nanostructures extend slightly below the average substrate level as depicted in Fig. 4e. No additional faceting or Au diffusion into the substrate bulk is observed, as it was for the Au/InSb(001) system. The process of Au/In alloy formation has recently been studied by F. Wang et al.[41] for the Au catalyzed InP nanorods growth. They have found that the stoichiometry of the In-Au alloy formed depends on the diameter of nanorods due to In supersaturation in the Au nanoparticle and limited diffusion in the nanorods. In our case, however, we do not observe any nanostructure size dependent effects since the Au diffusion is not limited and we have an almost infinite InP surface.

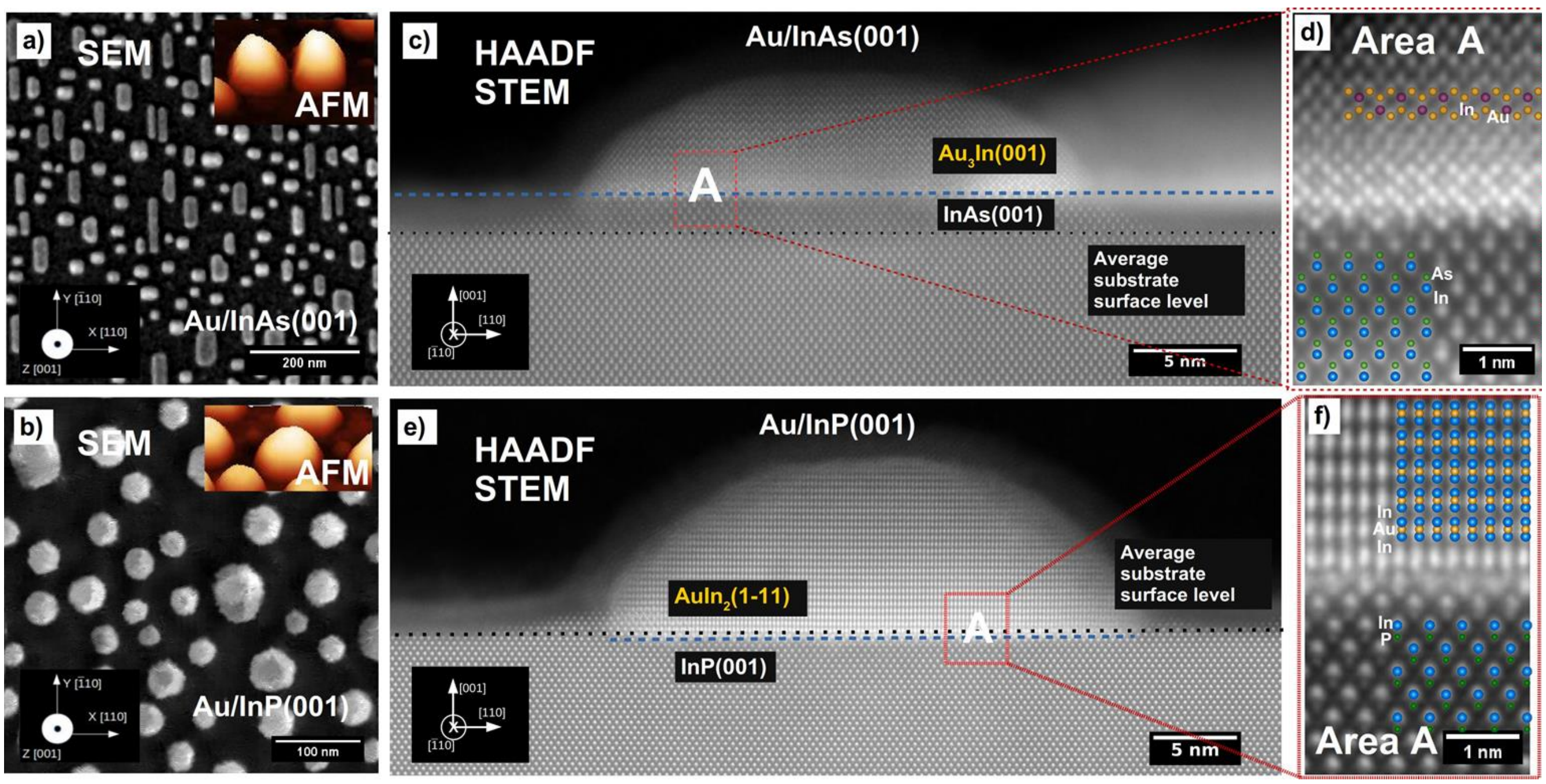

*Figure 4: Structural properties of the nanostructures grown on different Indium-based AIII-BV semiconductors after the deposition of 2ML of Au at 330ºC. The SEM and AFM morphologies of nanostructures on InAs(001) surface (in a) and on InP(001) surface (in b) are shown. In (c) an atomically resolved HAADF STEM image of cross-section of the nanostructure grown on InAs(001) together with detailed view on the $Au_3In$/InAs interface (in d) is presented. In (e) atomically resolved HAADF STEM image of cross-section of the nanostructures grown on InP(001) together with a detailed view on the $AuIn_2$/InP interface (in e) is shown. The atomic structural models of the phases are overlaid on the HAADF images in (d) and (f).*

## Ga-based AIII-BV semiconductors substrates - GaSb(001), GaAs(001) and GaP(001)

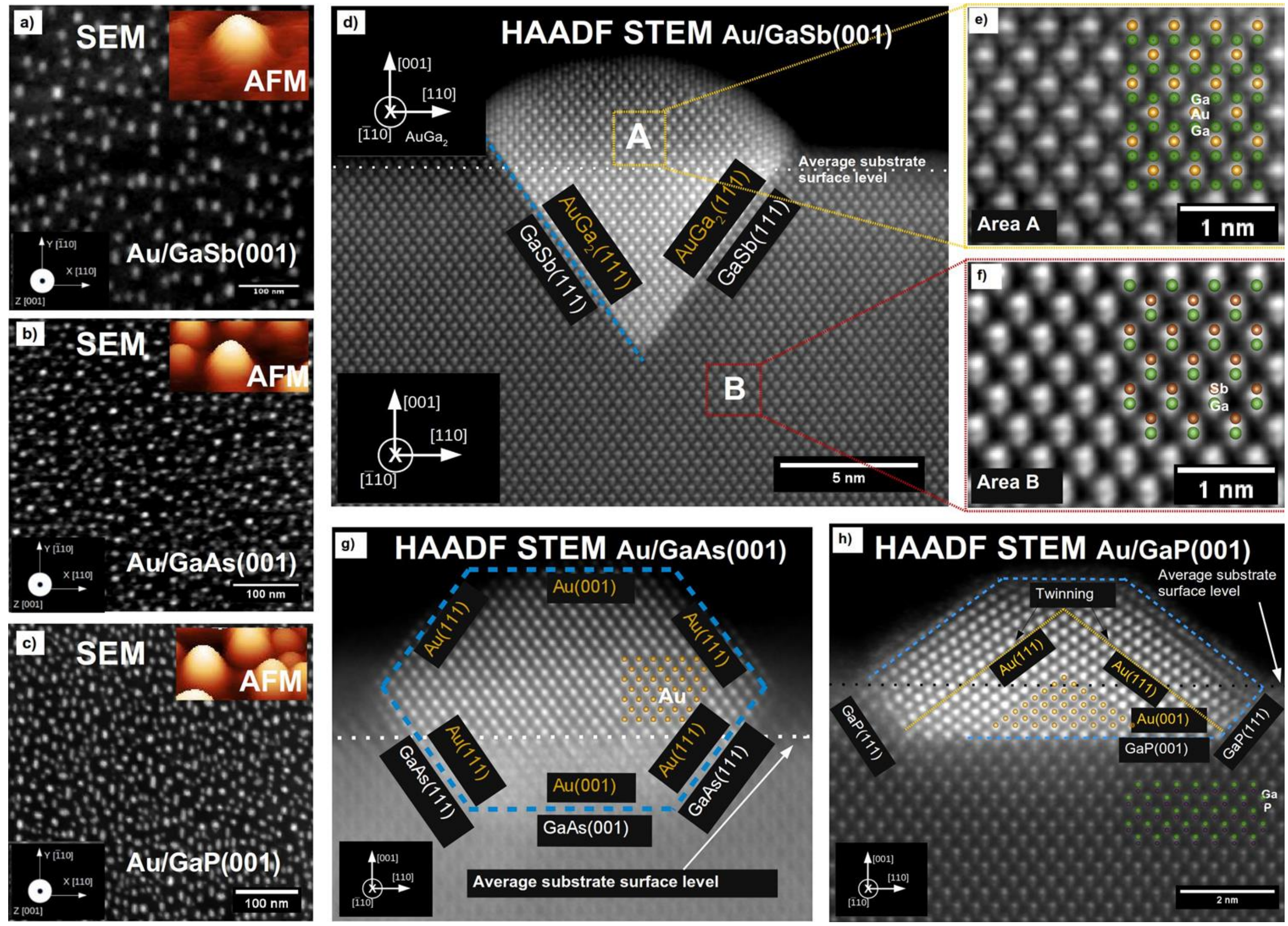

*Figure 5: The structural properties of the nanostructures grown on different Gallium-based AIII-BV semiconductors as a result of the deposition of 2ML of Au at 330ºC. The SEM and AFM morphologies of nanostructures grown on GaSb(001) (in a), GaAs(001) (in b) and on GaP(001) (in c) surfaces are presented. In (d) an atomically resolved HAADF STEM image of cross-section of the nanostructure grown on GaSb(001), with corresponding views of the GaSb substrate region (in e) and the AuGa2 nanostructure (in f), are shown. Atomically resolved HAADF STEM images of cross-section of the nanostructure grown on GaAs(001)(in g) and grown on GaP(001) (in h) are shown. The atomic structural models of the corresponding phases are overlaid on the HAADF images.*

The clusters/nanoislands formed after the deposition of 2ML of Au on the reconstructed c(3x1) GaSb(001) surface at 330 ºC were examined by SEM/AFM and the results are shown in Fig. 5a). The nanostructures are of an average lateral size of about 13 nm and of a constant height of 3 nm. In contrast to the Au/InSb, here no nanostructures' shape anisotropy is seen. In order to uncover the detailed internal nanostructures structure the atomically resolved HAADF STEM imaging of the nanostructure's cross section was performed as presented in Fig. 5d-f. By examining the interplanar spacing of the atomic columns, together with the results of the EDX experiments and the HAADF contrast, the nanostructure composition is identified as an $AuGa_2$ alloy (see Fig. 5e). No indications of Sb incorporation within the nanostructures have been found. Thus, the Au interactions with GaSb at elevated temperatures lead to the formation of one of the stable Au-Ga phases (see supporting information Fig. S2d and Table S2) and the releasing of Sb atoms into the vacuum. In this case, the process driving the chemical reaction would be $Au+2GaSb \rightarrow AuGa_2 + 2Sb \uparrow$ ($\Delta H=-0.046eV$).[33,34] We have also found that there is no Au dissolution in the GaSb bulk in the close vicinity of the nanostructure. It can be seen that the $AuGa_2$ nanostructures grow in registry with substrate with following epitaxial relationship: (001)$AuGa_2$//(001)GaSb and [110]$AuGa_2$//[110]GaSb. Such a relationship between these two phases was earlier already predicted based on the computational model.[34,35]

In the case of the Au/GaP(001) system, nanostructures with an average lateral size of about 6 nm, i.e., two times smaller than in the case of Au/GaSb, are formed (see Fig. 5c). In Fig. 5h the atomically resolved HAADF STEM image of the Au/GaP cross section is shown. The analysis of the interplanar spacing of the atomic columns indicates that the nanostructures formed are of the pure Au. The process driving the chemical reaction on this substrate surface is in this case: $5.9Au+GaP \rightarrow 5.9Au+Ga+P \uparrow$ ($\Delta H=0.823eV$).[33,34] For this reaction, we have estimated the number of Au atoms involved as a fraction of the nanostructure cross-section area which is below the average substrate surface. One can assume that only this part of the GaP underwent a reaction with Au. Otherwise, in the absence of Au-induced reaction, the nanostructures would grow fully on the sample surface. The Au interaction with GaP resulting in the formation of the AuGa alloy was earlier observed for the case of Au catalysed-nanorods growth with Au seeds (particles) of size lower than 15 nm.[16] In the present case, we do not observe any size-depend alloying effects, since the Au diffusion is not limited by the lateral substrate surface dimensions. The interfaces between the Au nanostructure and the substrate are of the following epitaxial relationship: (001)Au//(001)GaP and [110]Au//[110]GaP in agreement with the computational model predictions.[34,35]

For the third Ga-based substrate material studied, i.e., GaAs(001) the thermally induced self-assembly of 2 ML of Au leads to the development of small, 8 nm clusters, uniformly distributed on the substrate surface (Fig. 5b). The HAADF and EDX measurements provide that the clusters are of pure Au. They grow with their (001) planes in epitaxy with (001) GaAs substrate surface planes and they exhibit side walls of (111) facets. The epitaxial relationship in this case is also (001)Au//(001)GaAs and [110]Au//[110]GaAs in agreement with previous studies.[42] The Au structures are partially submerged into the GaAs substrate similarly to the case of GaP(001). The process driving the chemical reaction is $2.5Au+GaAs \rightarrow 2.5Au+Ga+As\uparrow$ ($\Delta H$=0.701eV).[33,34] Here, as in the GaP case, the number of Au atoms involved in the process is estimated as a fraction of the whole nanostructure area which is below the average substrate surface level. More details on the dynamics of the metallic nanostructures grown after thin Au layer deposition on GaAs(001) surface can be found in recent paper by present authors in Ref.[85]

# Discussion

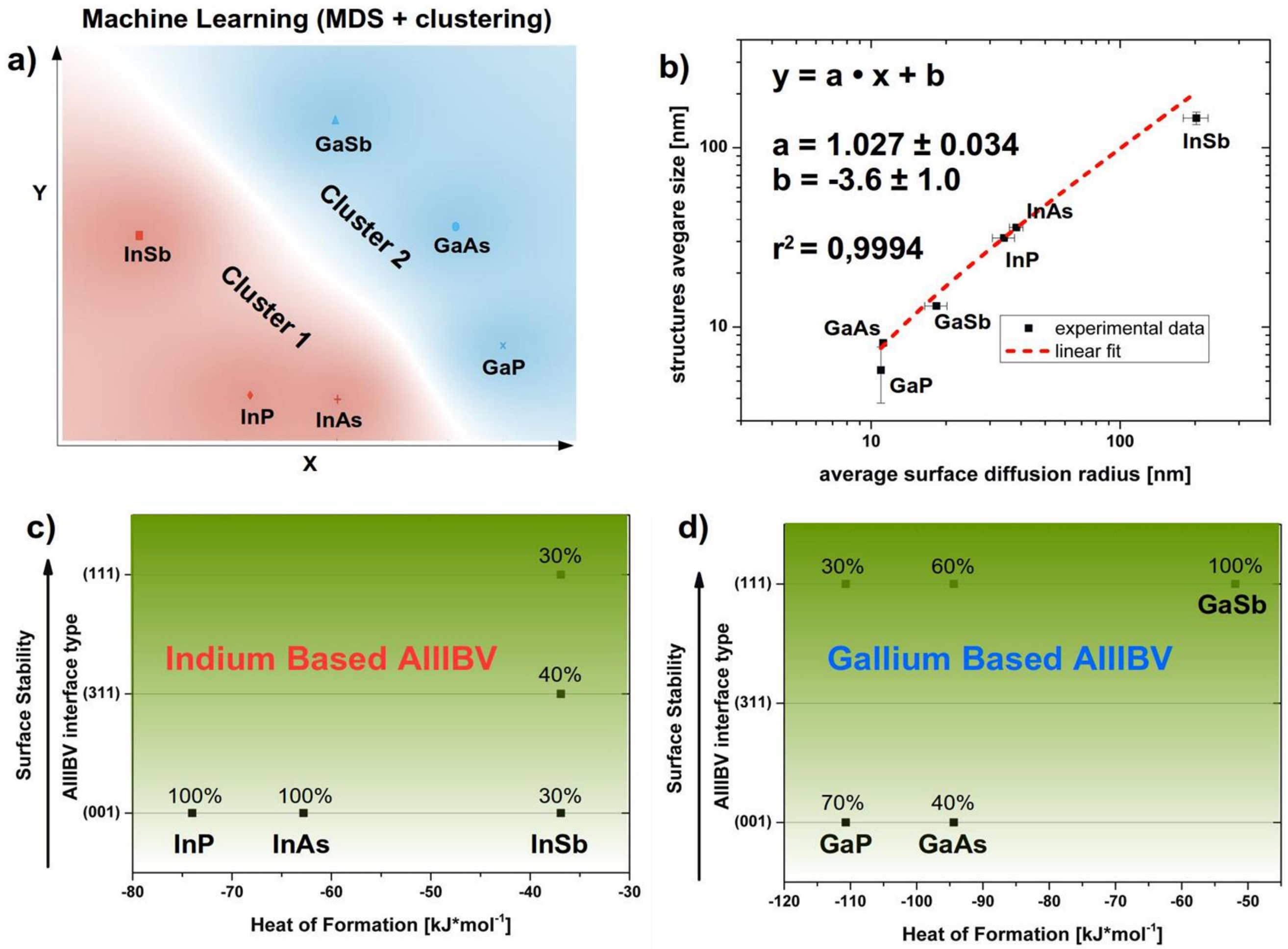

*Figure 6: Analysis of the properties of the nanostructures grown on InSb(001), InP(001), InAs(001), GaSb(001), GaAs(001), and GaP(001) surfaces after the deposition of 2ML of Au at 330 ºC. In (a) Machine Learning Multidimensional Scaling (MDS) 2D projection of the nanostructures multidimensional parameter space, together with the result of k-means clustering on the multidimensional data indicated by the colours red and blue. Two groups of points that are close together in the multidimensional space of nanostructures parameters were found. In (b) the dependence of the nanostructure's average lateral size on the average surface diffusion radius is shown. The red dashed line is a linear fit to the data points. A nanostructure-substrate interface type for indium- and gallium-based semiconductors versus the heat of formation (binding energy) for the AIII-BV semiconductor is presented in (c) and (d), respectively. The diagrams show AIII-BV crystallographic planes at the interface, ordered according to their stability (surface energy) and their relative occurrence.*

A set of the seven experimental parameters of the studied nanostructures derived from the SEM, AFM and TEM measurements (i.e., size, surface density, diffusion radius, height, nanostructure's volume under the substrate surface, Au content, and number of Au atoms needed to release one AIII metallic atom on the surface - Supporting Information Table S3) were analysed by Machine Learning methods (multidimensional scaling – MDS[44, 86]) to extract the existing relations between these parameters. This approach allows to visualize the multidimensional system on the 2D plot without any assumption on the particular relation between these parameters. To do so, we used the k-means clustering method.[45, 86] In Fig. 6a the computed MDS projection of the map into 2D plot together with the results of the k-means clustering is presented. The k-means clustering approach separate data for indium- and gallium-based AIII-BV substrates pointing out that Ga-based and In-based AIII-BV semiconductors interact differently with Au. This is related to the different AIII-BV compound stability when in contact with Au.[16] Furthermore, the morphological parameters, like the average size, surface density, and the average surface diffusion radius of the nanostructures grown split into separate two groups when they are studied as a function of the heat of formation HoF (binding energy) of the AIII-BV semiconductor.[46] Again those two groups are related to indium- and gallium-based substrates (see Fig. S4 in supporting information).

There is also a strong linear correlation between the structure average size and the average surface diffusion radius for all the studied AIII-BV substrates as depicted in Fig. 6b. This indicates that independently of the chemical interaction mechanism between Au and AIII-BV semiconductors, the nanostructures are mainly formed as a results of the surface diffusion of adatoms and their nucleation. In Fig. 6 c-d the supporting AIII-BV facets formed at the interfaces with the nanostructures are plotted as a function of the heat of formation for AIII-BV semiconductors. The relative occurrence of the particular facet is given and the facets are ordered accordingly to their stability, i.e., surface energy.[47] In Fig. 6c it is seen that for the InSb, the semiconductor with the lowest binding energy, there is high Au-induced etching of the (001) surface, and facets with a higher stability (311) and (111) are formed. As the semiconductor binding energy increases (like the ones of InAs and InP) the efficiency of Au-induced etching of the (001) surface is meaningfully decreases, for the same experimental conditions, thus only (001) facets are present. The situation is similar for gallium-based semiconductors as shown in Fig. 6d. For the lowest binding energy (GaSb) the (001) surface etching is very high, so only (111) facets are formed. When the binding energy increases (for GaAs) the (001) facets start to appear (40%) and with still the increasing of binding energy (like the one for GaP) the (001) facet starts to dominate (70%). Such hierarchy in the surface etching by Au atoms and facets formation is also observed for single-component semiconductors such as Si or Ge.[48,49]

The above mentioned observations point out that mechanism of Au-induced breaking of AIII-BV bonds is the main process of the nanostructure interface formation due to the preferential etching of the substrate surface, leading finally to the formation of the facets having the smallest surface energies.[47] Thus, the metallic nanostructures grow by forming preferentially (111) crystallographic facets which are of the lower surface energy.[50,51]

In Table 1 the experimentally evaluated number of Au atoms engaged in the release of one AIII metallic atom together with a leading chemical reaction and heat of formation of an AIII-BV semiconductor are presented.

| **Examined System AIII-BV Heat of Formation [kJ*mol$^{-1}$]**[46] | **Chemical Reaction (ΔH - calculated reaction enthalpy**[33,34] **)** | **Number of Au atoms needed to release one AIII metallic atom on the surface (this experiment)** |
|---|---|---|
| **2ML Au on InSb(001)** -36.9 | Au+2InSb → $AuIn_2$ + 2Sb ↑ (ΔH= − 0.215eV) | **0.5** |
| **2ML Au on InAs(001)** -62.8 | 3Au+InAs → $InAu_3$+As ↑ (ΔH= − 0.007eV) | **3** |
| **2ML Au on InP(001)** -74 | Au+2InP → $AuIn_2$ + 2$P$ ↑ (ΔH=0.172eV) | **0.5** |
| **2ML Au on GaSb(001)** -51.9 | Au+2GaSb → $AuGa_2$ + 2Sb ↑ (ΔH= − 0.046eV) | **0.5** |
| **2ML Au on GaAs(001)** -94.4 | 2.5Au+GaAs → 2.5Au+Ga+As ↑ (ΔH=0.701eV) | **2.5(0.4)** |
| **2ML Au on GaP(001)** -110.7 | 5.9Au+GaP → 5.9Au+Ga+P ↑ (ΔH=0.823eV) | **5.9(0.34)** |

*Table 1: The chemical reactions contributing to the Au nanostructures formation on AIII-BV semiconductor surfaces resulting from 2ML Au deposition at 330$^{o}$C, as derived from our measurements.*

We have found that depending on the HoF of the studied AIII-BV semiconductors the metallic nanostructures grown can be of pure Au or Au-AIII alloy. This is related to the chemical substrate stability against interaction with Au. When the system is strongly bound (more negative heat of formation), it is difficult to supply for the AIII metal (In, Ga) on the surface to form an alloy, so the formed structures phase is comprised of pure Au.

## Conclusions

In conclusion, we have investigated the bases of Au interactions with AIII-BV semiconductors at the atomic level via a thermally induced self-assembly process of 2 ML of Au with different AIII-BV (001) semiconductor surfaces. It has been found that the Au-rich nanostructures grown in this process are crystalline and in epitaxy with the AIII-BV substrate. The Au diffusion into the bulk lattice is very efficient only for the InSb crystal which is material of the lowest binding energy in the group of studied AIII-BV semiconductors.To rationalize the mechanism of Au diffusion into In-Sb lattice the DFT calculations were performed to determine the most stable atomic arrangement of diffusing Au atoms into the In-Sb lattice. Furthermore, a difference has been found in the Au interactions with In-based and Ga-based semiconductors, as confirmed by Machine Learning MDS and k-means analysis. The analysis uncovered also that independently of the strength of Au interactions with AIII-BV semiconductor, the main driving force of nanostructures formation is the surface diffusion and nucleation of adatoms produced by the chemical reactions induced by Au atoms. Our studies also show that the chemical composition of the nanostructures depend on the AIII-BV semiconductor binding energy (heat of formation): as the AIII-BV binding energy increases the efficiency of Au-induced etching of the substrate surface decreases so for the AIII-BV of highest binding energy the nanostructures are made of pure gold. Our fundamental studies which have shed new light on the interactions between Au and the AIII-BV semiconductors at the atomic level, as presented in our thoroughly systematic work, will enhance the understanding of AIII-BV nanowires growth and will improve the construction of any future semiconductor electronic devices based on AIII-BV technology.

# Methods

*Au/AIII-BV sample preparation*

N-doped epi-ready indium antimonide – InSb(001), indium arsenide – InAs(001), indium phosphide – InP(001), gallium antimonide – GaSb(001), gallium arsenide – GaAs(001) and gallium phosphide - GaP(001) crystals were mounted on molybdenum plates. Prior to the introduction into the UHV Molecular Beam Epitaxy (MBE) system, with a base pressure of $1 \cdot 10^{-10}$ mbar, the samples were rinsed with isopropanol and ethanol and finally dried in a flow of air. The sample temperature was measured by a pyrometer (LumaSense , model IGA 140) with $\varepsilon=0.55$ for InSb and $\varepsilon=0.5$ for InAs, InP, GaSb, GaAs and GaP. The substrate surfaces were initially out-gassed for 1h at 150 °C. In order to remove the oxide layer the samples were exposed to low energy, 700 eV $Ar^+$ ion bombardment at a 60 deg incident angle at room temperature (RT). Subsequently, the substrate surfaces were cleaned in cycles of ion-beam irradiation at T=400 °C (InSb), T=427 °C (InAs), T=450 °C (InP), T=440 °C (GaSb) T=500 °C (GaAs), T=520 °C (GaP) and annealing at T=450 °C (InSb), T=477 °C (InAs), T=470 °C (InP), T=500 °C (GaSb), T=550 °C (GaAs), T=570 °C (GaP) until the c(8x2) InSb(001), (4x2) InAs(001), (4x2) InP(001), (3x1) GaSb(001), c(8x2) GaAs(001), (4x2) GaP(001) reflection high energy diffraction (RHEED) pattern was observed (see supporting information Fig. S3 and Fig. 1a). Such a cleaning procedure results in atomically flat surfaces as already imaged with an atomic force microscopy.[52] On the prepared surfaces 2 ML of Au was deposited at sample temperature of 330 °C and a rate of 0.1 ML $min^{-1}$ as checked by a quartz microbalance. After Au deposition the samples were cooled down to RT at a rate of 10 C $min^{-1}$.

*Au/AIII-BV sample characterization and analysis*

The samples were transferred in an ambient condition to a Dual Beam SEM/FIB FEI Quanta 3D FEG microscope equipped with an EDAX EDX/EBSD/WDS microanalysis system for further investigation; this being installed at the Jagiellonian University's Institute of Physics, Krakow Poland. The size and surface density of the developed nanostructures were evaluated from SEM images using free software ImageJ/FIJI .[53] The nanostructure size was calculated as the square root of the nanostructure area. The average diffusion radius r i.e., the average distance between the nanostructures, was calculated from the nanostructures surface density D accordingly to $r=1/2\frac{1}{\sqrt{D}}$. The extracted results were presented in dependence of studied AIII-BV semiconductors ordered accordingly to their heat of formation (binding energy),[46] where the most negative value corresponds to the strongest bound system. The surface morphologies of the samples were also imaged by the 5500 Agilent Atomic Force Microscope in a tapping mode to provide the height of the nanostructures. Later, the atomically resolved HAADF STEM measurements where the contrast is proportional to the atomic number Z and to the sample thickness,[54] were performed using a FEI (S)TEM Titan3 G2 60-300 microscope equipped with a monochromator, a probe Cs corrector (DCOR) and ChemiSTEM technology (an X-FEG field-emission electron gun and four windowless detectors of the Super-X EDX system)[55] operated at 300 kV and installed at AGH, Krakow Poland. HAADF-STEM images were acquired with a convergence angle of 20 mrad and a probe current of 80 pA. The EDX chemical composition maps were collected using a FEI Tecnai Osiris 200 kV TEM microscope equipped with ChemiSTEM technology installed at the Jagiellonian University, Krakow, Poland. For the chemical composition determination the Cliff-Lorimer method was used, as implemented in the ESPRIT manufacturer software from Bruker. For the TEM measurements thin foils (lamellae) were prepared by a SEM/FIB dual beam system (FEI Quanta 3D FEG) using a focused ion beam (FIB).

The identification of the formed phases in the examined systems, as verified by the calculated phase diagrams from the Materials Project[33–35] and OQMD[56,57] as well as experimental ones[17,19–21] (see supporting information Fig. S2, Table S2), was based on the analysis of atomically resolved HAADF STEM measurements by examining the interplanar spacing and angles together with the results of the performed EDX experiments (see supporting information Table S4) and the HAADF contrast. The simulations of the identified phases structural models, within the proper zone axis projection, as prepared by free VESTA software,[58] were overlaid onto the HAADF experimental data.

The determination of the strain fields was performed by means of Geometric Phase Analysis (GPA)[59] of atomically resolved HAADF STEM images using the free software Strain++.[60] The ellipticity of the atomic columns was extracted from atomically resolved HAADF STEM images by means of the free software Atomap.[40]

The multivariate statistical analysis of nanostructures parameters (height, size, density, chemical composition etc., see Table S3 in the supporting information) was performed by Machine Learning Multidimensional Scaling (MDS)[44, 86] and k-means clustering[45, 86] together with Silhouettes scoring[61] for the number of clusters estimation using the free software Orange.[62]

The Multidimensional Scaling (MDS)[44, 86] is a technique which finds a two dimensional projection of the multidimensional data by trying to preserve the distance between the points as much as possible. So, the points with similar properties in multidimensional space are close together in the MDS projection. This method takes as input dissimilarities between pairs of items in a matrix form and outputs matrix of coordinates (simply speaking this is like giving as the input the distances between the cities and as the outputs the coordinate of each city). Using this technique one can visualize multidimensional set of data as a simple X-Y scatter plot. Here, the points for the nanostructures with similar physical properties (like height, size, density) are close together (for details about MDS and its applications please look into Ref. [63,64]).

The k-means clustering[45, 86] is a method which searches for the group of points that are close together in the multidimensional space (clusters). It assumes that each point is closer to its own cluster center than to different cluster centers. This is accomplished by minimizing the sum of squares between the points and cluster centers (centroids) defined as arithmetic mean of all points which belong to the cluster. The cluster search is performed iteratively (repetitively) to optimize in each iteration the cluster centers. The initial cluster centers are selected randomly. The conditions for the successful clustering, is the stability of the cluster center positions in the next iteration (for details about k-means please look into Ref.[65]).

*Machine Learning HAADF STEM Image Quantification*

The HAADF STEM image quantification was performed for the Au/InSb system to distinguish between the columns containing Au atoms and the pure In-Sb columns (assuming a constant sample thickness). The image was segmented into cells containing atomic columns[66,67]. This has the advantage that all the image scattering is associated with a particular atomic column. Intermixing between the Au and In-Sb phases is studied column-by-column, in a similar way to Ref.[49] As a result of the noise inherent to the experiment, we used Machine Learning algorithms i.e., Random Forest,[68,69] which uses a multiple of decision trees which are trained on the subset of the data to classify (group into classes) the full dataset. The classified data comes with a derived classification probability. The random forest could be simply understood as a "wisdom of the crowd", in a computer one creates a virtual crowd (decision trees) which are trained and can answer questions (classify data) i.e., this part of the data is class A and the other one is B (for details about Random Forest and its applications please look into Ref.[70,71]).

The Random Forest method was used as implemented in Trainable Weka Segmentation[37] to statistically distinguish between these phases, as successfully used in HAADF STEM Tomography.[72] Here, we used experimental HAADF signal references for training of pure Au columns (from AuIn2 nanowire) and pure In-Sb columns (from the bulk), the following training features for cells containing atomic columns were used: mean, median, variance, maximum, minimum. The full dataset i.e., atomic columns were classified into In-Sb columns and Au columns together with classification probability for Au and In-Sb columns. The resulted probability of finding Au is directly proportional to the number of Au atoms in the atomic column (here assuming a constant sample thickness) and since $Probability_{Au}+Probability_{InSb}=1$, so the probability of finding Au is directly the Au atoms concentration in the sample (value "1" corresponds to the concentration of 100 atomic %), for details see supporting information (Fig. S1, Table S1). The estimated relative uncertainty of the performed HAADF STEM chemical quantification as a result of the local sample thickness variation is 10.2%. The estimated maximal relative uncertainty of the performed quantification related to the validity of the whole chemical quantification procedure is 12.6% (in comparison with EDX measurements). For details see the supporting information on pages S3-S4.

*Density Functional Theory (DFT) calculations*

The quantum-chemical calculations of gold build up into the InSb lattice were performed by using DFT/GGA with the use of the VASP[73–75] code with the following settings: the energy cut-off was set to 400 eV, the sampling of the irreducible Brillouin zone was done according to the Monkhorst-Pack[76] scheme with irreducible Brillouin zone IBZ sampling in the range of 0.008-0.03 $Å^{-1}$; employed was the Blöchl's projector augmented wave (PAW)[77,78] method for representing valence-core interactions together with the PBE[79] functional. The dispersion effects were accounted for by the empirical Tkatchenko-Scheffler correction.[80] The Methfessel-Paxton[81] smearing with $\sigma$ width of 0.1 eV was used. For solving the Kohn-Sham equations the SCF convergence criterion was an energy change between two successive iterations lower than $10^{-5}$ eV. Geometry optimization parameters were selected in order to obtain corrections to the forces acting on ions of less than 0.001 eV/Å. The cell optimisation was performed *via* the Birch-Murnaghan equation of state.[82] The model cell was built as the supercell of 1×8×1 (for higher Au concentrations) to 4×8×4 (for lower Au concentrations) primitive InSb unit cells. The thermodynamic functions were taken from the harmonic approximation approach and were verified by the classical Born-Oppenheimer molecular dynamics (MD) based calculations of Gibbs free energy. The MD calculations were performed using Universal Force Field (UFF) with 1 fs time step (sufficient for the systems with heavy atoms and for the temperature used), the total time of simulations was equal to 500 ps. The NPT ensemble with a Nosé[83] thermostat (T = 330 ° barostat (p = 0.0 GPa) were used in the simulations.

## Author contributions

B.R.J. and A.J. contributed to the characterization of the samples by RHEED/SEM and RHEED/SEM/TEM data analysis and interpretation. A.J. prepared the samples in UHV. A.K., G.C., A.K. and A.C-F. contributed to the HAADF STEM measurements. P.I. contributed to the TEM and EDX measurements. B.R.J. contributed to the FIB sample preparation and to the HAADF STEM image quantification by Machine Learning, the ellipticity analysis of the atomic columns and the Machine Learning multivariate statistical analysis of the data. W.P. contributed to the DFT and MD simulation of Au build up into an InSb lattice. K.S. contributed to the AFM measurements. B.R.J. prepared the manuscript together with A.J. and F.K. in consultation with all the authors. F.K. initiated and organized this project.

## Acknowledgments

We gratefully acknowledge the contribution of R. Abdank Kozubski for his fruitful discussions.
The support of the Polish National Science Center (UMO-2015/19/B/ST5/01841) and of the Polish Ministry of Science and Higher Education under the grant 7150/E-338/M/2016 is acknowledged.

## Competing financial interests

The authors declare no competing financial interests.

# Supporting Information

## Towards Understanding of Gold Interaction with AIII-BV Semiconductors at Atomic Level

B.R. Jany[a*], A. Janas[a], W. Piskorz[c], K. Szajna[a], A. Kryshtal[b], G. Cempura[b], P. Indyka[c,d], A. Kruk[b], A. Czyrska-Filemonowicz[b], F. Krok[a]

[a]The Marian Smoluchowski Institute of Physics, Jagiellonian University, Lojasiewicza 11, 30-348 Krakow, Poland
[b]The International Centre of Electron Microscopy for Materials Science, AGH University of Science and Technology, 30-059 Krakow, Poland
[c]The Faculty of Chemistry, Jagiellonian University, ul. Gronostajowa 2, 30-387 Krakow, Poland
[d]Malopolska Centre of Biotechnology (MCB), Jagiellonian University, Krakow, 30-387, Poland

*Corresponding author e-mail: benedykt.jany@uj.edu.pl

# Machine Learning HAADF STEM image quantification

The HAADF STEM imaging mode provides structural images where intensities are proportional to both the thickness and mean atomic number Z[1] (Au columns appear brighter than In-Sb columns). By assuming a constant sample thickness, this can be efficiently used to distinguish between columns containing Au atoms and pure In-Sb columns. The image was segmented into cells containing atomic columns[2,3]. This has the advantage that all image scattering is associated to some atomic column. Due to the noise inherent to the experiment, intermixing between the Au and In-Sb phases is studied atomic column-by-column, similar as in[4]. Here we used Machine Learning algorithms (Random Forest) as implemented in Trainable Weka Segmentation[5] to statistically distinguish between these phases, as successively used in HAADF STEM Tomography[6]. For the Trainable Weka Segmentation the following training features were used: mean, median, variance, maximum, minimum. The classes were balanced. The rest settings were set on their default values (classifier: fast random forest of 200 trees with 2 features per tree). The reference areas were used for the Au atomic columns and In-Sb atomic columns, as indicated in Fig. S1a. As a result of image quantification of analysis area (area with Au atom diffusion into the bulk InSb crystal -Fig. S1a) the Au probability map and InSb probability map is computed Fig. S1b-c. Since the probability of finding Au is directly proportional to the number of Au atoms in the atomic row (assuming constant sample thickness) and $Probability_{Au}+Probability_{InSb}=1$, so the probability of finding Au is directly the Au atoms concentration in the sample (value “1” corresponds to the concentration of 100 atomic %). The histogram of Au probability (Au atomic concentration) from the area with Au atom diffusion into the bulk InSb crystal is presented in Fig. S1d. Several local maxima are seen Fig. S1d and Table S1. This is compared with EDX measurements of this region as presented in histogram of Au atomic concentration from EDX Fig. 1e and Table S1. The quantitative HAADF STEM shows more local maxima in comparison to the EDX measurements, so in this case in more locally sensitive to the changes of Au atomic concentrations.

| **HAADF STEM quantification**<br>Au atomic concentrations<br>[X at.] | **EDX measurements**<br>Au atomic concentration<br>[X at.] |
|---|---|
| 0.037 | – |
| 0.075 | 0.068 |
| 0.12 | – |
| 0.23 | – |
| 0.28 | – |

*Table S1: Au atomic concentaration in the* area with Au atom diffusion into the bulk InSb crystal as measured by HAADF STEM image quantification and EDX. In this case the quantitative HAADF measurements are more sensitive to local Au concentration then EDX, since more local maxima are visible (value “1” corresponds to the concentration of 100 atomic %).

To check the validity of whole chemical quantification procedure i.e. how it estimates the true unknown Au concentration, we compared the results of HAADF STEM quantification with EDX measurements. Assuming that EDX concentration (0.068) is a mixture of two components as seen by HAADF quantification (0.037 and 0.075), which gives central value of 0.060, one can estimate the maximal relative quantification uncertainty as relative difference (0.068-0.060)/0.060*100% which is equal to 12.6%.

In Fig. S1f  Au probability map (Au atomic concentration) from Area1 is presented, one can see in details that different amount of Au atoms are built up into the In-Sb lattice atomic positions.

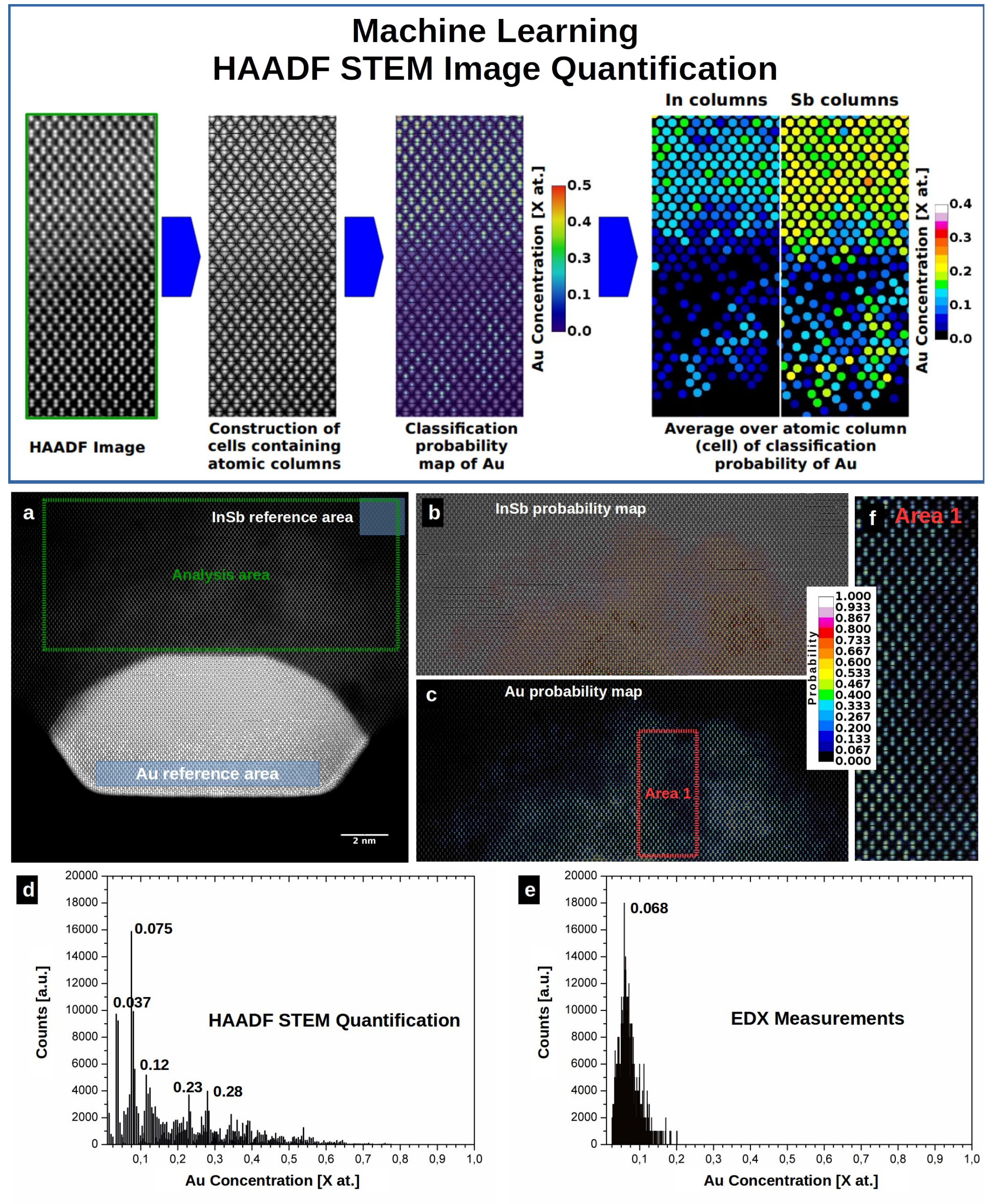

*Fig. S1: HAADF STEM image quantification of Au/InSb at 330C. Graphical presentation of the idea of the method (top). a) atomically resolved HAADF STEM image, Au and InSb reference area marked, b) quantification result, InSb probability map overlaid on HAADF image from analysis area, c) quantification result, Au probability map overlaid on HAADF image from analysis area, d) histogram of Au probability from analysis area. Since the probability of finding Au is directly proportional to the number of Au atoms in the atomic row (assuming constant sample thickness) and ProbabilityAu+ProbabilityInSb=1, so the probability of finding Au is directly the Au atoms concentration in the sample. e) histogram of Au atomic concentration, as measured by EDX. The obtained Au probability (Au atomic concentration) from HAADF STEM quantification matches well the EDX measurements. The main maxima, corresponding to the main atomic concentrations are marked. It is seen that HAADF STEM is more locally sensitive than EDX since additional small maxima are visible. f) Au probability map (Au atomic concentration) from Area1, one can see in details that different amount of Au atoms are built up into the In-Sb lattice atomic positions. Value "1" corresponds to the concentration of 100 atomic %.*

**Local sample thickness variations changes for the HAADF quantification area of the thin foil sample prepared by FIB were estimated, and their influence on the chemical quantification result.**

The line profile 1 (Fig. S1-1b) was extracted from HAADF STEM image (Fig. S1-1a), from the area without Au diffusion. Since there is only one material in this area (namely InSb) so the changes of the HAADF Intensity in this area are only due to the local sample thickness changes, as seen in the line profile. Taking the histogram of the intensities from the linear profile Fig. S1-1c, one can calculate the mean HAADF intensity and the standard deviation of it. So the relative HAADF signal deviation related to the local sample thickness changes is equal to 0.28%.

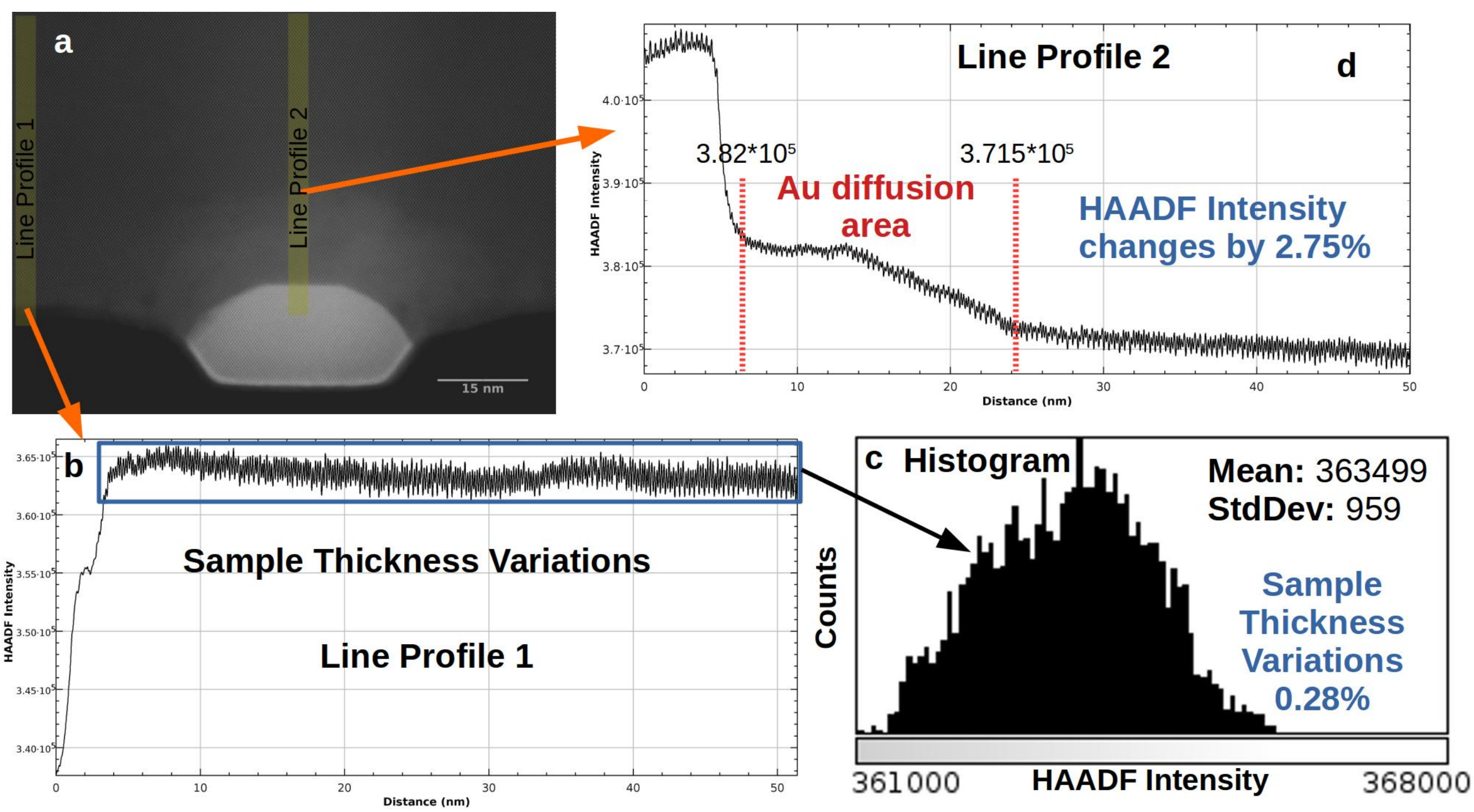

*Fig. S1-1: Local sample thickness variation changes for the HAADF quantification area of the thin foil sample of Au/InSb at 330C. a) HAADF STEM image of the analysis area, two line profiles indicated. b) Line profile 1 from HAADF STEM image a) from sample area without Au diffusion, the local HAADF signal changes correspond to local sample thickness variations. c) Histogram from Line profile 1 b) from the marked by blue rectangle area on the profile. It is seen that the distribution standard deviation (value of 958) is equal to 0.28% of the mean value (363499). So the relative HAADF signal deviation related to the local sample thickness changes is equal to 0.28%. d) Line profile 2 from HAADF STEM image a) through the nanowire and Au diffusion area. It is seen that the HAADF signal changes on the Au diffusion area by 2.75% (changes from $3.82*10^5$ to $3.715*10^5$) are above deviations related to the estimated local sample thickness changes 0.28%.*

One can now compare this values with the HAADF signal changes below the nanowire in the Au diffusion area, see Fig. S1-1d. It is seen that here the HAADF signal changes by 2.75% (changes from $3.82*10^5$ to $3.715*10^5$), which is related to Au diffusion into InSb. This HAADF signal changes related to the Au diffusion are above deviations related to the estimated local sample thickness changes. Next, one can estimate the influence of the local sample thickness changes on the HAADF chemical quantification performed by comparing the estimated values 0.28%/2.75%*100%=10.2%. So the estimated relative uncertainty of the performed HAADF STEM chemical quantification due to the local sample thickness variation is 10.2%.

# Au-AIII-BV Phase Diagrams

The Au-AIII-BV phase diagrams were calculated from First Principles using the generalized gradient approximation (GGA) approximation to density functional theory (DFT) and the DFT+U extension to it[7,8] by the Materials Project[9] and also OQMD[10,11]. In agreement with experimental phase diagrams[12–14].

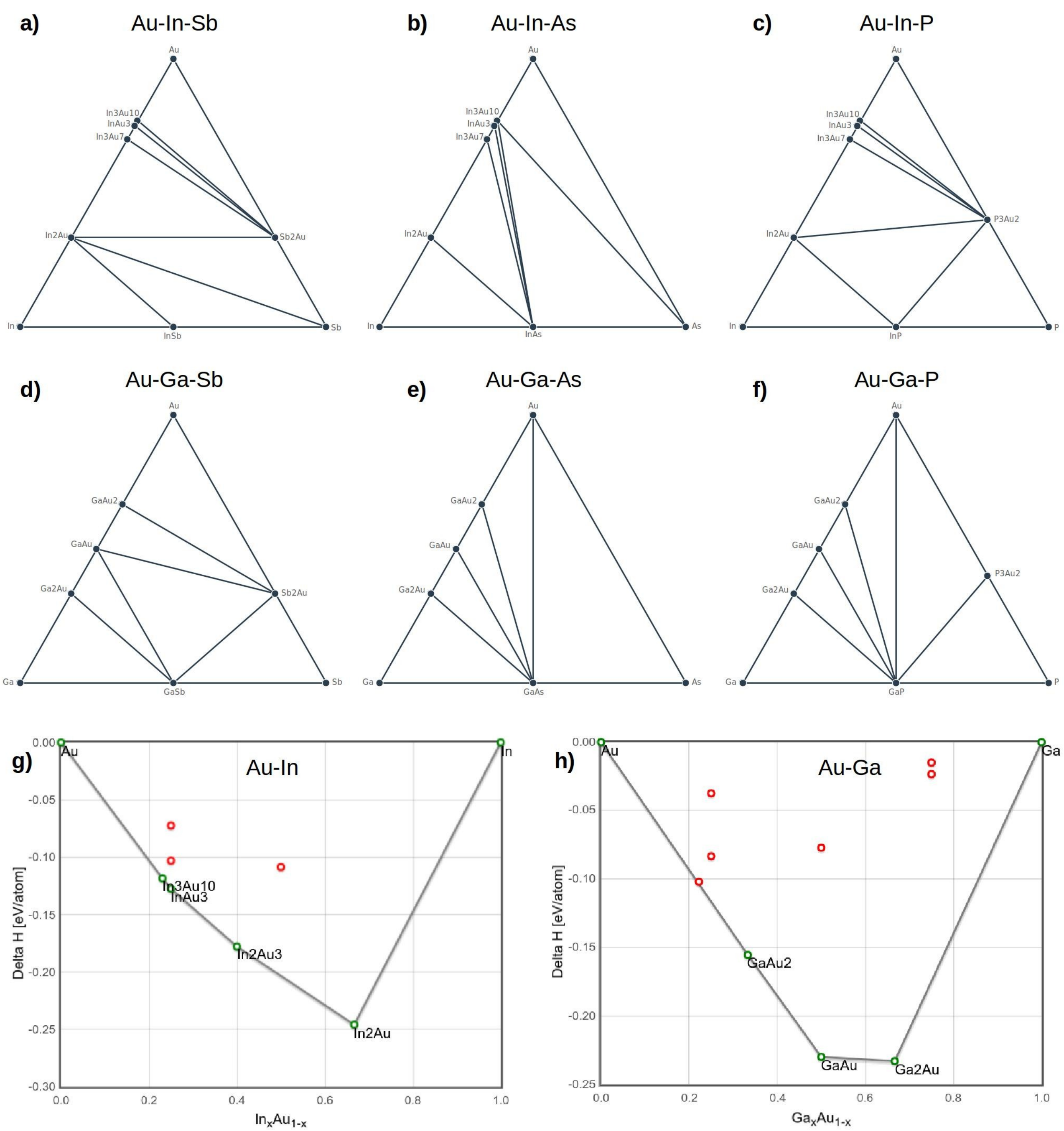

*Fig. S2: Theoretically calculated Phase Diagrams for Au-AIII-BV systems: a) Au-In-Sb, b) Au-In-As, c) Au-In-P, d) Au-Ga-Sb, e) Au-Ga-As, f) Au-Ga-P, g) Au-In, h)Au-Ga by the Materials Project[9] and also OQMD[10].*

| **Au-In System** | | **Au-Ga System** | |
|---|---|---|---|
| Phase | Formation Energy [eV] | Phase | Formation Energy [eV] |
| $AuIn_2$ | -0.246 | $AuGa_2$ | -0.234 |
| $Au_3In$ | -0.126 | AuGa | -0.227 |
| $Au_7In_3$ | -0.141 | $Au_2Ga$ | -0.151 |
| $Au_{10}In_3$ | -0.116 | $Au_7Ga_2$ | -0.102 |
| AuIn | -0.108 | $Au_3Ga$ | -0.083 |
| $AuIn_3$ | 0.016 | $AuGa_3$ | -0.023 |

*Table S2: Theoretically calculated Phases in the Au-In and Au-Ga system together with their formation energies by the Materials Project*[9] *and also OQMD*[10]*. Unstable phases also included. In agreement with experimental phase diagrams*[15,16]*.*

# RHEED patterns of the atomically clean and reconstructed AIII-BV surfaces

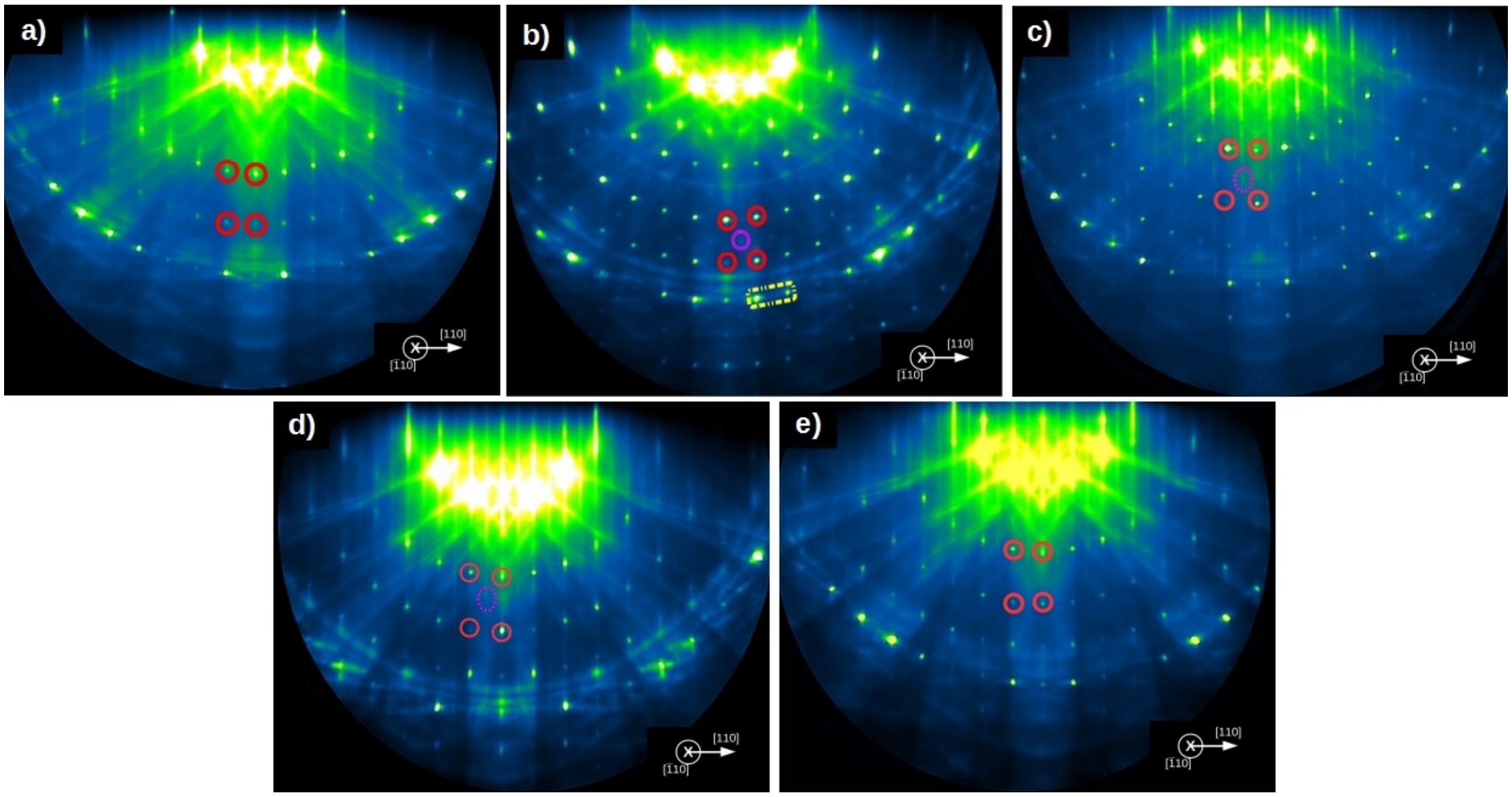

*Fig. S3: RHEED patterns of the atomically clean and reconstructed surfaces of a) (3x1) GaSb(001), b) c(8x2) GaAs(001), c) (4x2) GaP(001), d) (4x2) InAs(001), e) (4x2) InP(001). Main reconstruction spots marked.*

# Multivariate Statistical Analysis

The obtained from SEM, AFM and TEM measurements data on nanostructures formed in the Au/AIIIBV systems are presented in Table S3, as in detailed described in the main article. It is seen that for each AIIIBV system a set of seven parameters is used to describe it, forming seven dimensional parameter space. This is later used for multivariate statistical analysis using machine learning Multidimensional Scaling (MDS)[17] and k-means clustering[18].

| AIIIBV system | Average size [nm] | Surface density [1/um$^2$] | Surface diffusion radius [nm] | Average height [nm] | Nanostructure percent under the sample surface [%] | Au concentration in the nanostructure [atomic %] | Number of Au atoms needed to release one AIII metallic atom on the surface |
|---|---|---|---|---|---|---|---|
| **InSb** | 146.3 | 6.1 | 202.3 | 8.0 | 35 | 33.3 | 0.5 |
| **InAs** | 35.9 | 170.5 | 38.3 | 7.9 | 0 | 75.0 | 3 |
| **InP** | 31.4 | 214.4 | 34.1 | 11.3 | 6.45 | 33.3 | 0.5 |
| **GaSb** | 13.1 | 747.4 | 18.3 | 0.9 | 69 | 33.3 | 0.5 |
| **GaAs** | 8.2 | 2003.8 | 11.2 | 1.1 | 41.8 | 100.0 | 2.5 |
| **GaP** | 5.8 | 2088.5 | 10.9 | 1.4 | 13.4 | 100.0 | 5.9 |

*Table S3: Measured seven parameters of the formed nanostructures in the Au/AIIIBV systems.*

# Formed Nanostructures Properties

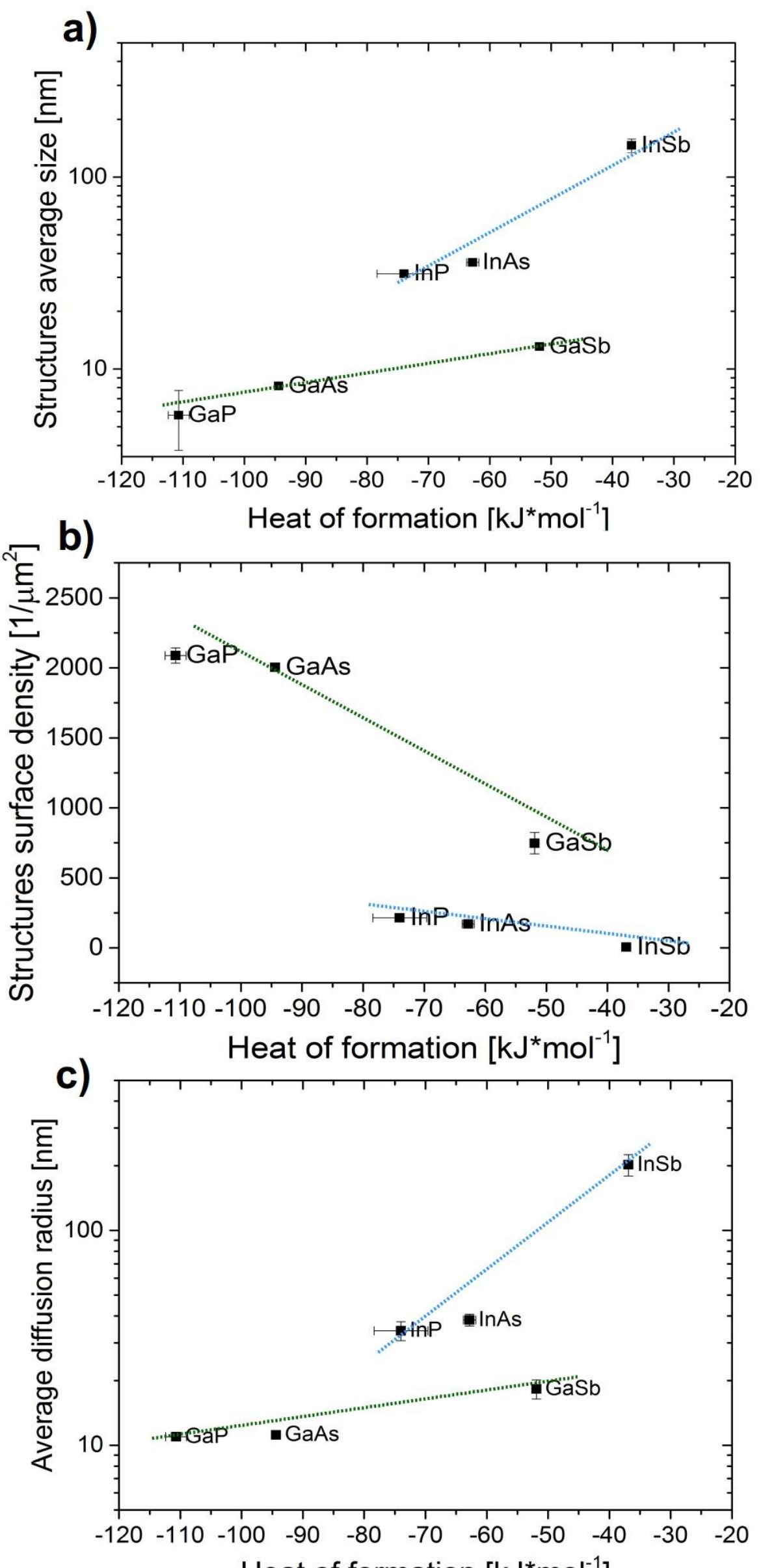

*Fig. S4: Nanostructures resulted from 2ML Au deposited on InSb(001), GaSb(001), GaAs(001), InP(001) and InAs(001) surfaces at 330C. Average sizes b), surface densities c) and average diffusion radius d) as a function of the heat of formation of AIII-BV semiconductor. The lines are drawn to guide the eye. Different behavior is observed for the structures grown on In and Ga rich surfaces.*

# Results of STEM EDX Analysis

For the examined Au/AIII-BV systems the STEM EDX measurements were performed in form of hyperspectral maps. To obtain EDX signal coming only from formed nanostructures the Machine Learning Blind Source Separation (BSS) analysis using Non Negative Matrix Factorization (NMF) was performed on hyperspectral EDX data accordingly to [19,20]. Later the chemical compositions were quantified by Cliff-Lorimer method, the results of quantification are presented in Table S4.

| Au/AIII-BV System | Results of EDX of bulk AIII-BV [at. %] | Results of EDX of Nanostructures [at. %] |
|---|---|---|
| 2ML Au on InSb(001) | In: 51.0(1.0) Sb: 49.0(1.0) | Au: 34.0(1.0) In: 66.0(1.0) |
| 2ML Au on InAs(001) | In: 51.3(1.3) As: 48.7(1.3) | Au: 73.3(1.3) In: 26.7(1.3) |
| 2ML Au on InP(001) | In: 47.3(2.7) P:52.7(2.7) | Au: 28.6(2.7) In: 71.4(2.7) |
| 2ML Au on GaSb(001) | Ga: 54.1(4.1) Sb:45.9(4.1) | Au: 37.1(4.1) Ga: 62.9(4.1) |
| 2ML Au on GaAs(001) | Ga: 47.9(2.1) As: 52.1(2.1) | Pure Au |
| 2ML Au on GaP(001) | Ga: 52.7(2.7) P: 47.1(2.7) | Pure Au |

*Table S4: Results of STEM EDX analysis for the Au/AIII-BV systems.*